\documentclass[a4paper,11pt]{article}

\usepackage[english]{babel}
\usepackage[T1]{fontenc}
\usepackage{authblk}
\usepackage{ucs}
\usepackage[utf8x]{inputenc}
\usepackage{amsmath}
\usepackage{amssymb}
\usepackage{amsfonts}
\usepackage{setspace}
\usepackage{graphicx}
\usepackage{nomencl}
\usepackage{subfigure}
\usepackage{color}
\usepackage{refstyle}
\usepackage{cleveref}
\usepackage{atbegshi}% http://ctan.org/pkg/atbegshi
\AtBeginDocument{\AtBeginShipoutNext{\AtBeginShipoutDiscard}}

\date{}

\begin{document}

\title{Multi-chimera states in the Leaky Integrate-and-Fire model}

\author[1]{
    N.D. Tsigkri-DeSmedt}%\inst{1}%\thanks{Designed and implemented the class style}
\author
    [1,2]{J. Hizanidis}%\inst{1,2}%\thanks{Did numerous tests and provided a lot of suggestions}
\author
    [3,4]{P. H{\"o}vel}%\inst{3,4}
\author
   [1,*]{A. Provata}\thanks{Corresponding Author: a.provata@inn.demokritos.gr}\\
%}

% Institutes for affiliations are also joined by \and,
\affil[1]{\it Institute of Nanoscience and Nanotechnology, National Center for 
      Scientific Research ``Demokritos'' Athens, Greece}
  %\email{a.provata@inn.demokritos.gr; n.tsigkri@inn.demokritos.gr}}
\affil[2]{\it Crete Center for Quantum Complexity and Nanotechnology, Department of Physics,
   University of Crete
   Heraklion, Crete, Greece.}
   %\email{hizanidis@physics.uoc.gr}}\\
\affil[3]{\it Institut f{\"u}r Theoretische Physik, Technische Universit{\"a}t Berlin,
   Berlin, Germany}
\affil[4]{\it Bernstein Center for Computational Neuroscience Berlin, Humboldt-Universit{\"a}t zu Berlin,
   Berlin, Germany}
   %\email{phoevel@physik.tu-berlin.de}}\\ 

\maketitle

\begin{abstract}
We study the dynamics of identical leaky integrate-and-fire neurons with symmetric non-local coupling. 
Upon varying control parameters (coupling strength, coupling range, refractory period) we investigate the system's 
behaviour and highlight the formation of chimera states. We show that the introduction of a refractory period enlarges 
the parameter region where chimera states appear and 
affects the chimera multiplicity.\end{abstract}

%------------------------------------------------------------------------------
\section{Introduction}
\label{sect:introduction}

The study of the dynamics and in particular collective behaviour of coupled oscillators has received great interest from scientists in 
different fields varying from chemical and mechanical systems to neuroscience and beyond
\cite{PIK01}. A very interesting 
and unexpected synchronisation phenomenon that was first observed in identical coupled oscillators 
is the so-called \textit{chimera state}. This is a dynamical scenario in which part of the oscillators are 
synchronised, while simultaneously others are not synchronised. These states were first observed 
in 2002 by Kuramoto and Battogtokh \cite{KUR02a}, while the term ``chimera'' was coined later, 
in 2004, by Abrams and Strogatz \cite{ABR04}. 
Potential applications of chimera states include 
the unihemispheric sleep that appears in dolphins and some birds, which sleep with one eye open
meaning that half of the brain is synchronised and half is not synchronised, power grids and 
social systems \cite{Panaggio}. On one hand, this surprising phenomenon has been observed numerically in 
various neuron models such as leaky integrate-and-fire, Kuramoto phase oscillators, Hindmarsh-Rose, 
FitzHugh-Nagumo, and SNIPER/SNIC model \cite{Abrams,Abrams2,BRU, Hizanidis,KUR02a,OME13,OME15,VUE14a}.
On the other hand, experimental verifications \cite{HAG12,LAR13,MAR13,TIN12,WIC13} do not include examples from neuroscience so far. 
This gives rise to an even greater interest to study chimera states as it may 
lead to a better understanding of information processing in neuron networks. 
In this study we examine the effect of different control 
parameters on the appearance of chimera states for Leaky Integrate-and-Fire (LIF) neuronal oscillators
% Our main focus is the effect of a resting time that we call \textit{refractory period}. 
% We consider coupled LIF neurons 
that are arranged in a 1-dimensional regular ring topology. 
We compare the behaviour of coupled LIF units with and without refractory period and we find 
that in both cases chimera states appear. We show that when the refractory period is 
introduced the chimera states are enhanced and their multiplicity increase.

In the next section we introduce the single and coupled LIF models. Subsections 2.2 and 2.3 
describe the coupled LIF model with and without a refractory period, respectively. In Sec. 3 we show 
the development of chimera states in a network of coupled LIF neurons. We demonstrate the differences 
in the form of chimera states between a network of coupled LIF neurons with and without a refractory 
period. Finally, the main conclusions are recapitulated in Sec. 4.

%------------------------------------------------------------------------------
\section{The leaky integrate-and-fire model}

\subsection{The single neuron model}
\label{subsec:single}
\begin{figure}[ht!]
\begin{center}
\includegraphics[clip,width=0.4\linewidth]{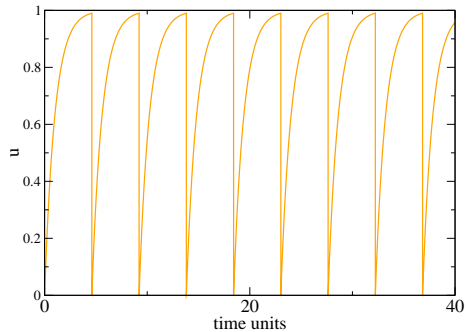}
\caption{\label{fig1} (Colour online) Dynamic evolution of a single neuron in time according to Eq.~\ref{eq:single}. Parameters: $\mu=1$, $u_{\text{th}}=0.99$, $u_{\text{rest}}=0$.
}
\end{center}
\end{figure}

The LIF model is a simple model for spiking neurons \cite{Ermentrout} which was introduced in 
1907 by Louis Lapicque. It describes the dynamical evolution of the membrane potential of a single 
neuron. Figure~\ref{fig1} depicts the spiking behaviour of the membrane potential of a single LIF neuron in time.

The membrane potential $u(t)$ evolves according to the following equation
\begin{eqnarray}\label{eq:single}
% \begin{split}
\dot{u}(t)=-u(t)+\mu
% \end{split}
\end{eqnarray}
with a reset condition
\begin{eqnarray}
\forall \> u(t)=u_{\text{th}} \> \Rightarrow \> \lim_{\varepsilon \to 0}u(t+\epsilon)=u_{\text{rest}}
\end{eqnarray}
where $u_{\text{th}}$ is the threshold of the potential and $\mu>u_{\text{th}}$ denotes a 
constant. In LIF, whenever the membrane potential 
reaches the threshold $u(t)= u_{\text{th}}$, a spike is fired and the membrane potential is 
instantaneously reset to the rest state $u_{\text{rest}}$. In this study the potential in the rest state is set equal to zero, $u_{\text{rest}}=0$. 

\subsection{Non-locally coupled LIF neurons}

\begin{figure}[ht!]
\begin{center}
\includegraphics[clip,width=0.4\linewidth]{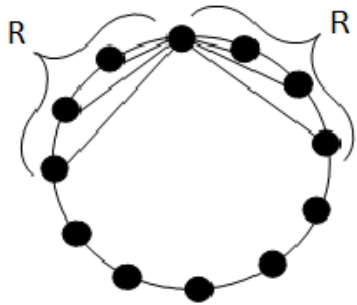}
\caption{\label{fig2} Topology the considered one-dimensional ring network.
}
\end{center}
\end{figure}

Neurons ``fire'' electrical signals as a result of receiving inputs from other neurons. This observation sets the need  of studying a network of coupled neurons.
We study a network of $N$ LIF neurons that are arranged in a regular ring topology with non-local connections, that is, each element is coupled to $R$ nearest neighbours on either side, as schematically depicted in Fig.~\ref{fig2}. The dynamic evolution in time of this system is determined by 
\begin{eqnarray}\label{eq:coupled}
\dot{u_i}(t) &= -u_i(t)+\mu+\frac{\sigma}{2R}\sum\limits_{j=i-R}^{i+R}[u_i(t)-u_j(t)]
\end{eqnarray}
with the same reset mechanism for each element as described in Sec.~\ref{subsec:single}. Here, $\sigma$ is the coupling strength and $R$ denotes the coupling range. The index $i$ has to be taken modulo $N$. The network nodes are considered identical, that is, they have the same system and coupling parameters.
%  $\mu$ and $u_{\text{th}}$ have the same physical meaning as in Eq (1).

The study of a system of coupled oscillators \cite{Vilela,Lindner,Orhan} 
involves the identification of parameter regions where synchronisation occurs. 
% Relevant 
% parameters in this system are the coupling strength $\sigma$, the coupling range $R$, the 
% topology and as will be discussed in section 2.3, the refractory period. 
In the next 
sections, we investigate the effect of the coupling strength $\sigma$ and the coupling range $R$ on synchronisation phenomena with special focus on chimera states.

\subsection{Coupled neurons with a refractory period}

\begin{figure}[ht!]
\begin{center}
\includegraphics[clip,width=0.4\linewidth]{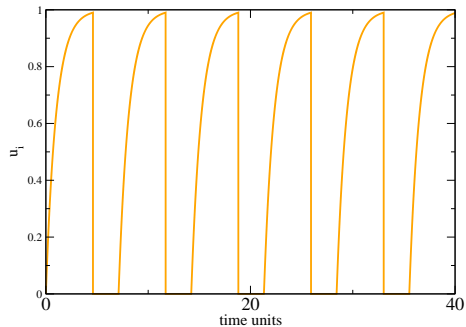}
\caption{\label{fig3} (Colour online) Dynamic evolution of a LIF neuron with a refractory period of $p_r=1$. Other parameters as in Fig.~\ref{fig1}.
}
\end{center}
\end{figure}

In many neuron models, the neuron stays in its rest state for a certain period 
of time after firing. In order to take this into account, we consider a refractory period $p_r$ \cite{Kouvaris}. 
The refractory period is a time interval, during which a neuron remains at rest after firing and is not able to trigger an additional spike.

The dynamics of the refractory LIF model is described by the equations of the coupled LIF neurons
system Eq.~\ref{eq:coupled}, except that after firing each neuron remains at the rest state for time $p_r$.
Figure~\ref{fig3} depicts the spiking behaviour of the membrane potential of a single LIF neuron with a refractory period $p_r=1$.

%------------------------------------------------------------------------------
\section{Chimera states in coupled leaky integrate-and-fire neurons}

We investigate the appearance of chimera states of a network of LIF neurons with and without refractory period. 
In references \cite{Panaggio,Tattini,Luciolii,Zare,Olmi} 
chimera states appear in the LIF system, for different realisations of the model and of the coupling geometry. 
In reference \cite{Olmi} the authors have shown the existence of chimera states in coupled LIF systems with delay dynamics. In this study, 
we show that the presence of a refractory period favours their appearance, while 
at the same time has an effect on their multiplicity. 
The refractory period is different from delayed self-feedback in the sense 
that the former introduces a dead (resting) time after firing while in the latter each neuron receives input not only by its 
neighbours but also by its past states.

\subsection{Without refractory period}

\begin{figure}[ht!]
\begin{center}
\includegraphics[clip,width=0.7\linewidth]{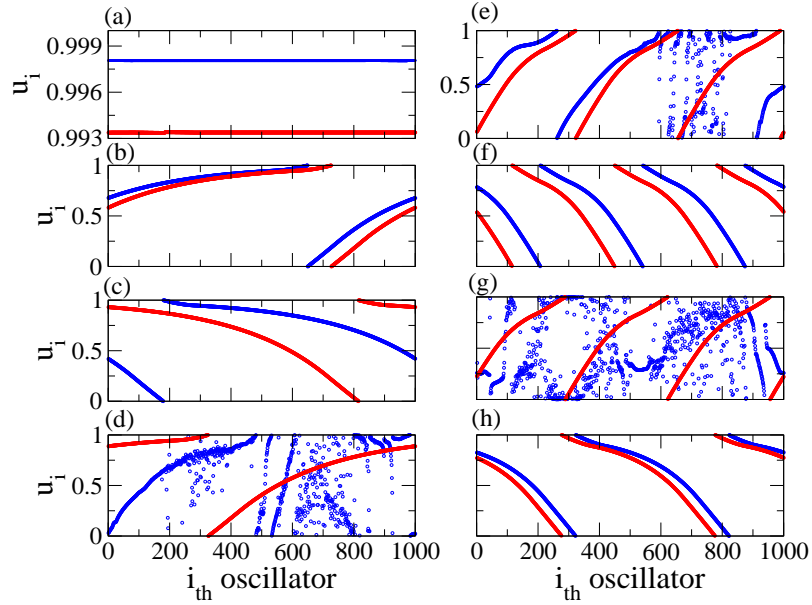}
\caption{\label{fig4} (Colour online) Snapshot of the membrane 
potential $u_i$ for different values of the coupling strength: (a) $\sigma=0.4$, 
(b) $\sigma=0.52$, (c) $\sigma=0.54$, 
(d) $\sigma=0.56$, (e) $\sigma=0.565$, (f) $\sigma=0.57$, (e) $\sigma=58$, (g) $\sigma=0.6$. 
The blue line corresponds to $t=1000$ time units 
and the red line to $t=9000$ time units. Other parameters: $N=1000$, $u_{\text{th}}=0.98$, $R=100$ and $\mu=0.99$. 
}
\end{center}
\end{figure}

\begin{figure}[ht!]
\begin{center}
\includegraphics[clip,width=0.6\linewidth]{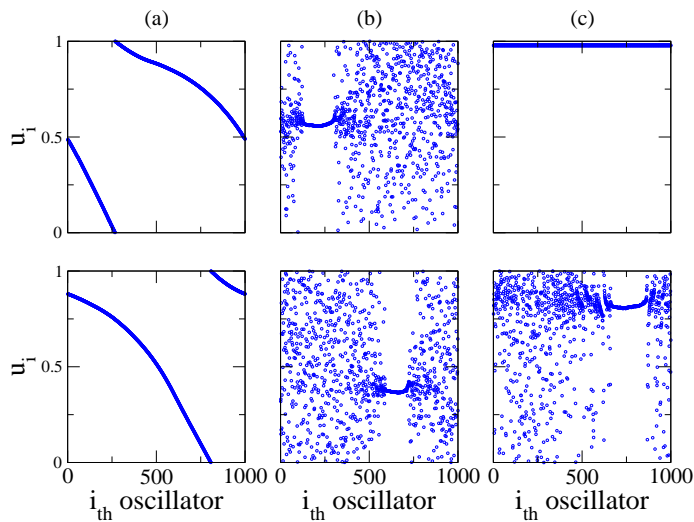}
\caption{\label{fig5} (Colour online) Snapshots of the membrane potential $u_i$ for different values of the coupling parameters $R$ and $\sigma$.
The upper panel corresponds to $\sigma=0.565$, while the lower one to $\sigma=0.7$. The coupling range is (a) $R=200$, (b) $R=300$ and (c) $R=400$. 
Other parameters as in Fig~\ref{fig4}.
}
\end{center}
\end{figure}

\begin{figure}[ht!]
\begin{center}
\includegraphics[clip,width=0.7\linewidth]{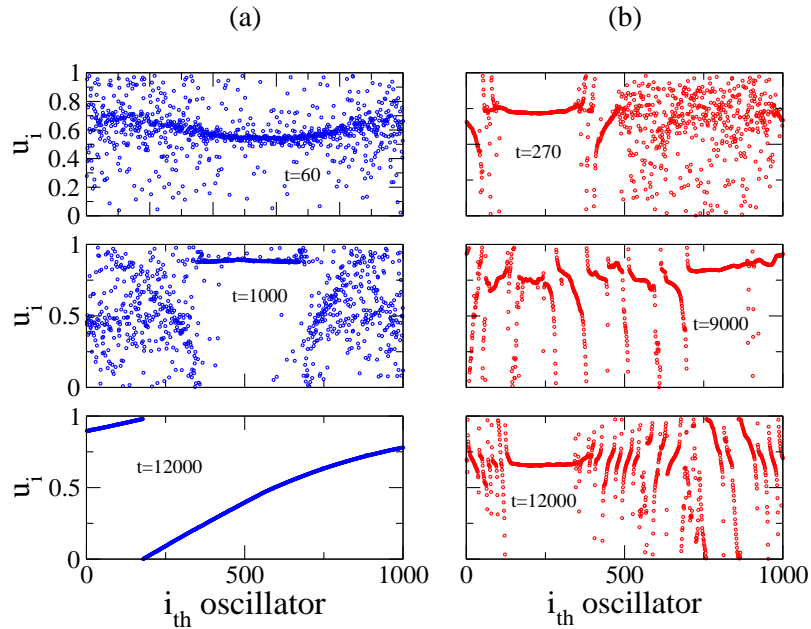}
\caption{\label{fig6} (Colour online) Snapshots of the membrane potential $u_i$ 
for different time units and for different initial conditions in panels (a) and (b). Parameters: $N=1000$, $u_{\text{th}}=0.98$, $\sigma=0.565$, $R=350$ and $\mu=1$.
}
\end{center}
\end{figure}

In the following, starting from random initial conditions $u_i,\> \{ i=1,...,N \}$ distributed over
the interval [0,1], we investigate the appearance of chimera states for a finite network 
of $N=1000$ neurons by varying the coupling strength $\sigma$ and the coupling range $R$. See Fig.~\ref{fig4}.
We observe that for $R=100$ chimera states do not appear for very small values of the 
coupling strength $\sigma \leq 0.5$ nor for $\sigma \geq 0.6$. They are found for 
intermediate values of the coupling strength, such as $\sigma=0.565$, as shown in 
Fig.~\ref{fig4}(e). Notice that the chimera states 
observed in this case are transient and disappear for longer times. See red curves in Fig.~\ref{fig4}.
% This doesn't mean that they definitely will not appear for higher 
% or lower values of the coupling strength, but if they do they will appear for different values of the coupling range. 

The appearance of chimera states at a certain value of the coupling strength also 
depends on the value of the coupling range $R$. More specifically, 
we show that as we increase the coupling range $R$, chimera states appear for a larger
value of $\sigma$, as shown in Fig.~\ref{fig5}. 
Thus the range of the values of the coupling strength that favour the appearance 
of chimera states, shifts following the change of the coupling range. 

Chimera states are highly dependent on initial conditions. Figure~\ref{fig6} 
shows the temporal evolution of chimera states starting from two different 
random initial conditions in columns (a) and (b), respectively. All other parameters are the same. 
We find that when the system starts from an initial state (a) it reaches 
complete synchronisation, while when it starts from initial state (b) 
a chimera state is formed as shown in the plots. 

\newpage
\subsection{With refractory period}

The study of the network of coupled LIF neurons shows that this network displays the phenomenon of chimera states which are mostly transients. 
We now examine the effect of the refractory period in their spatial form and temporal evolution. Using values of the coupling strength $\sigma$ 
and the coupling range $R$ for which we observed chimera states in the original coupled LIF system, we now consider the influence of a non-zero refractory period $p_r$.

\begin{figure}[ht!]
\begin{center}
\includegraphics[clip,width=0.75\linewidth]{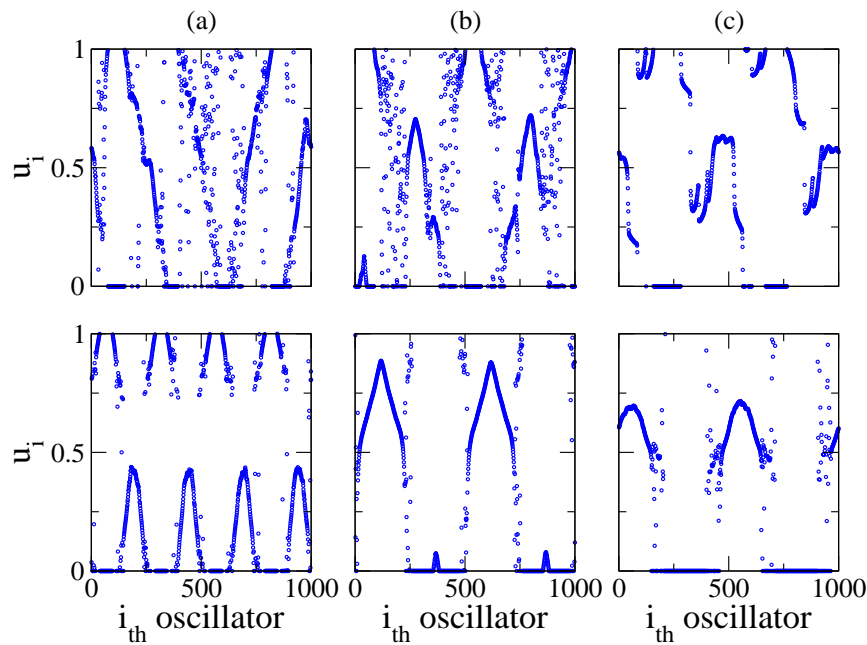}
\caption{\label{fig7} (Colour online) Snapshots of the membrane potential $u_i$ for different values of the coupling parameters $R$ and $p_r$. The upper panel corresponds to $p_r=500$ time units and the lower panel corresponds to $p_r=1000$. The coupling range is (a) $R=200$, (b) $R=300$ and (c) $R=400$. Other parameters are $N=1000$, $u_{\text{th}}=0.98$,
$\sigma=0.565$, $\mu=0.99$, $t=9000$ time units.
}
\end{center}
\end{figure}

In Fig.~\ref{fig7} we demonstrate that the refractory period enhances the appearance of chimera states. When varying the refractory period it is natural to use a time scale for comparison that is intrinsic to the system. As this reference, we use the period $T$ of the oscillations of the coupled LIF unit. 
\begin{figure}[ht!]
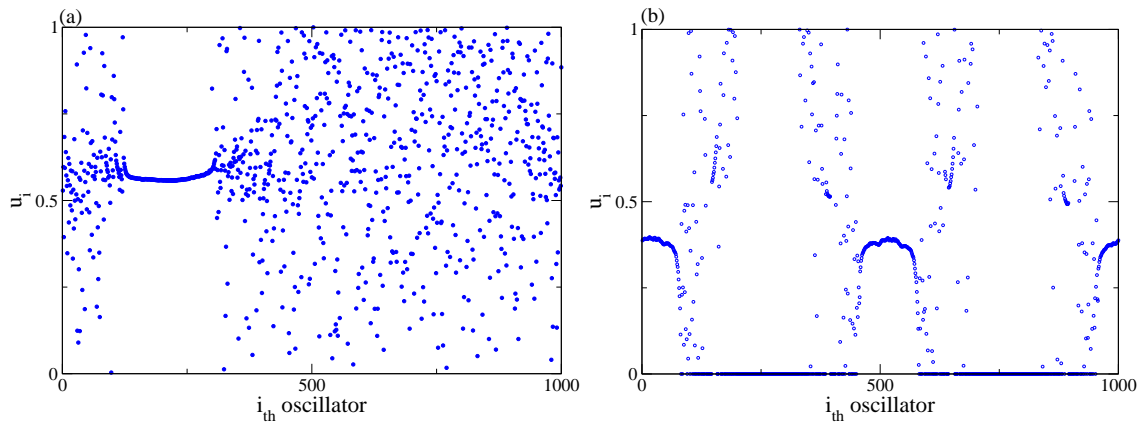

\includegraphics[clip,width=0.50\linewidth]{A0.565.eps}
\includegraphics[clip,width=0.48\linewidth]{n-1500.eps}
\caption{\label{fig8} (Colour online) Snapshots of the membrane potential 
$u_i$: panel (a) depicts the formation of chimera states in a network without a refractory period and 
panel (b) shows the formation of chimera states in a network with $p_r=0.5T$. Other parameters are: $N=1000$, 
$\sigma=0.565$, $u_{\text{th}}=0.98$ , $R=300$ and $\mu=0.99$.
}
\end{figure}

In Fig.~\ref{fig8} we compare the system of the $N=1000$ oscillators 
behaviour with and without a refractory period. 
For $p_r=0$, as show in Fig.~\ref{fig8}(a), the chimera state has one coherent and one incoherent region. 
On the contrary, in Fig.~\ref{fig8}(b) the chimera state that is formed in the system with 
$p_r=0.5T$ has four incoherent and four coherent regions. 

% We next demonstrate the results of a refractory period that takes 
% values between $0.1T\leq p_r \leq T$. We show that the refractory period favours 
% the appearance of  chimera states, which in this case survive throughout 
% the simulation. In addition we stress that when we introduce a refractory period 
% the multiplicity of chimera states changes. 

\begin{figure}[ht!]
\begin{center}
\includegraphics[clip,width=0.75\linewidth]{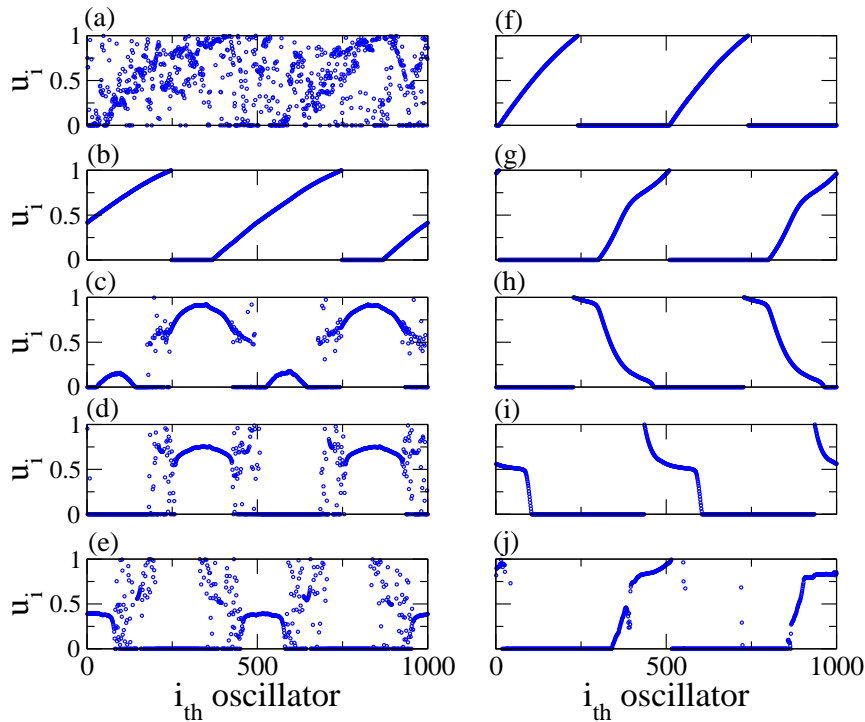}
\caption{\label{fig9} (Colour online) Snapshots of the membrane potential $u_i$ in space 
for different values of the refractory period $p_r$, (a) $p_r=0.1T$, (b) $p_r=0.2T$, (c) $p_r=0.3T$, (d) $p_r=0.4T$, 
(e) $p_r=0.5T$, (f) $p_r=0.6T$, (g) $p_r=0.7T$. (h) $p_r=0.8T$, (i) $p_r=0.9T$, (j) $p_r=T$. Other parameters are: $N=1000$, $u_{\text{th}}=0.98$, $R=300$ and $\mu=0.99$.
}
\end{center}
\end{figure}

In the following, we elaborate on the effect of the refractory period. 
As shown in Fig.~\ref{fig9} the chimera multiplicity changes as $p_r$ varies from $0.1T$ to $T$. 
Notice that the chimera states appear only for intermediate values of the refractory period and that the 
number of coherent and incoherent regions for fixed values of the coupling strength and coupling range 
remains constant. A potential interpretation of this behaviour is that the neurons by sections slightly 
differ in phase. 
Intuitively, the small but substantial values of the refractory period facilitate 
the grouping because the condition $u_i(t)=0$ forces neighbouring elements to synchronise locally in 
the rest state. The grouping of neurons in sections, influences the neurons on the boundaries between 
sections, which destabilise and become asynchronous. 

\section{Conclusions}
\label{conclusions}

Chimera states on a non-locally coupled network of LIF neurons highly depend on the 
combination of coupling strength, coupling range and refractory period $p_r$. The analysis of a network of $N=1000$ neurons has shown that 
chimera states appear for intermediate values of the coupling strength. 
Furthermore, the emergence of chimera states also depends on the value of the coupling range. 
More specifically, we have observed that as the coupling range increases, the range 
of coupling strengths that favour the appearance of chimera states shifts to higher values. 
% This range of values seems to be constant despite the shift. 
Additionally, we have noticed that a crucial control parameter for the occurrence of chimera states 
is the refractory period, a resting period between two consecutive excitations of a neuron. 
We have shown that the refractory period helps the chimera states survive for longer periods, while at the same time is responsible for the formation 
of multiple coherent and incoherent regions. The number of coherent and incoherent regions 
for fixed values of the coupling strength and the coupling range, does not depend on the value of the refractory 
period.
\par Our results represent only a first approach to the study of the effect of control parameters in a LIF network 
and to the phenomenon of partial synchronisation (more specifically chimera states).
Future work should address quantitative investigation of the parameter regions which favour chimera states and could include 
additional parameters related to the experimentally measured time-scales of biological neurons.

\section{Acknowledgements}
This work was supported by the German Academic Exchange Service (DAAD) and the Greek State Scholarship Foundation IKY
within the PPP-IKYDA framework. This research has been cofinanced by the European Union (European Social Fund–ESF) and Greek national 
funds through the Operational Program “Education and Lifelong Learning” of the National Strategic Reference Framework (NSRF) -- Research Funding Program: 
THALES. Investing in knowledge society through the European Social Fund. Funding was also provided by NINDS R01-40596. The research work was partially
supported by the European Union's Seventh Framework Program (FP7-REGPOT-2012-2013-1) under grant agreement n316165. PH
acknowledgez support by DFG in the framework of the Collaborative Research Center 910.

%------------------------------------------------------------------------------
% Refs:
%

\label{sect:bib}
%\bibliographystyle{plain}
%\bibliographystyle{alpha}
%\bibliographystyle{unsrt}
%\bibliographystyle{abbrv}
%\bibliography{YSC-TSIGKRI2015}

%------------------------------------------------------------------------------
%\section*{References}
% \bibliography{mybibfile,ref}
%\section*{References}
% \bibliography{mybibfile,ref}

\begin{thebibliography}{10}
\expandafter\ifx\csname url\endcsname\relax
  \def\url#1{\texttt{#1}}\fi
\expandafter\ifx\csname urlprefix\endcsname\relax\def\urlprefix{URL }\fi
\expandafter\ifx\csname href\endcsname\relax
  \def\href#1#2{#2} \def\path#1{#1}\fi

\bibitem{PIK01}
A.~Pikovsky, M.~G. Rosenblum, J.~Kurths, Synchronization, A Universal Concept
  in Nonlinear Sciences, Cambridge University Phys.Rev.Ess, 2001.

\bibitem{KUR02a}
Y.~Kuramoto, D.~Battogtokh, Coexistence of coherence and incoherence in
  nonlocally coupled phase oscillators, Nonlin. Phen. in Complex Sys. 5 (2002)
  380.

\bibitem{ABR04}
D.~M. Abrams, S.~H. Strogatz, Chimera states for coupled oscillators, PRL 93
  (2004) 174102.
\newblock \href {http://dx.doi.org/0.1103/PhysRevLett.93.174102}
  {\path{doi:0.1103/PhysRevLett.93.174102}}.

\bibitem{Panaggio}
M.~J. Pannagio, D.~Abrams, Chimera states: coexistence of coherence and
  incoherence in networks of coupled oscillators, Nonlinearity 28 (2015) R67.
\newblock \href {http://dx.doi.org/10.1088/0951-7715/28/3/R67}
  {\path{doi:10.1088/0951-7715/28/3/R67}}.

\bibitem{BRU}
N.~Brunel, M.~C.~W. Van~Rossum, Lapicque’s 1907 paper: from frogs to
  integrate-and-fire, Biol Cybern
\href {http://dx.doi.org/10.1007/s00422-007-0190-0}
  {\path{doi:10.1007/s00422-007-0190-0}}.

\bibitem{Hizanidis}
J.~Hizanidis, V.~Kanas, A.~A.~Bezerianos, T.~Bountis, Chimera states in
  networks of nonlocally coupled hindmash-rose neuron models, International
  Journal of Bifurcation and Chaos 24 (2013) 03.
\newblock \href {http://dx.doi.org/10.1142/S0218127414500308}
  {\path{doi:10.1142/S0218127414500308}}.

\bibitem{Abrams}
D.~Abrams, R.~R.~Mirollo, S.~H. Strogatz, D.~A. Wiley, Solvable model for
  chimera states of coupled oscillators, Phys.Rev.E 101 (2008) 084103.
\newblock \href {http://dx.doi.org/10.1103/PhysRevLett.101.084103}
  {\path{doi:10.1103/PhysRevLett.101.084103}}.

\bibitem{Abrams2}
D.~Abrams, S.~H. Strogatz, Chimera states in a ring of nonlocally coupled
  oscillators, International Journal of Bifurcation and Chaos 16 (2006) 21--37.
\newblock \href {http://dx.doi.org/10.1142/S0218127406014551}
  {\path{doi:10.1142/S0218127406014551}}.

\bibitem{OME13}
I.~Omelchenko, O.~E. Omel'chenko, P.~H{\"o}vel, E.~Sch{\"o}ll, When nonlocal
  coupling between oscillators becomes stronger: patched synchrony or
  multichimera states, Phys. Rev. Lett. 110 (2013) 224101.
\newblock \href {http://dx.doi.org/10.1103/physrevlett.110.224101}
  {\path{doi:10.1103/physrevlett.110.224101}}.

\bibitem{VUE14a}
A.~V{\"u}llings, J.~Hizanidis, I.~Omelchenko, P.~H{\"o}vel, Clustered chimera
  states in systems of type-{I} excitability, New J.~Phys. 16 (2014) 123039.

\bibitem{OME15}
I.~Omelchenko, A.~Provata, J.~Hizanidis, E.~Sch{\"o}ll, P.~H{\"o}vel,
  Robustness of chimera states for coupled {FitzHugh-Nagumo} oscillators, Phys.
  Rev. E 91 (2015) 022917.
\newblock \href {http://dx.doi.org/10.1103/physreve.91.022917}
  {\path{doi:10.1103/physreve.91.022917}}.

\bibitem{TIN12}
M.~R. Tinsley, S.~Nkomo, K.~Showalter, Chimera and phase cluster states in
  populations of coupled chemical oscillators, Nature Physics 8 (2012)
  662--665.
\newblock \href {http://dx.doi.org/10.1038/nphys2371}
  {\path{doi:10.1038/nphys2371}}.

\bibitem{HAG12}
A.~M. Hagerstrom, T.~E. Murphy, R.~Roy, P.~H{\"o}vel, I.~Omelchenko,
  E.~Sch{\"o}ll, Experimental observation of chimeras in coupled-map lattices,
  Nature Physics 8 (2012) 658--661.
\newblock \href {http://dx.doi.org/10.1038/nphys2372}
  {\path{doi:10.1038/nphys2372}}.

\bibitem{MAR13}
E.~A. Martens, S.~Thutupalli, A.~Fourri{\`e}re, O.~Hallatschek, Chimera states
  in mechanical oscillator networks, Proc. Nat. Acad. Sciences 110 (2013)
  10563.
\newblock \href {http://dx.doi.org/10.1073/pnas.1302880110}
  {\path{doi:10.1073/pnas.1302880110}}.

\bibitem{WIC13}
M.~Wickramasinghe, I.~Z. Kiss, Spatially organized dynamical states in chemical
  oscillator networks: Synchronization, dynamical differentiation, and chimera
  patterns, PLoS ONE 8~(11) (2013) e80586.
\newblock \href {http://dx.doi.org/doi:10.1371/journal.pone.0080586}
  {\path{doi:doi:10.1371/journal.pone.0080586}}.

\bibitem{LAR13}
L.~Larger, B.~Penkovsky, Y.~Maistrenko, Virtual chimera states for
  delayed-feedback systems, Phys. Rev. Lett. 111 (2013) 054103.
\newblock \href {http://dx.doi.org/10.1103/physrevlett.111.054103}
  {\path{doi:10.1103/physrevlett.111.054103}}.

\bibitem{Ermentrout}
B.~Ermentrout, Neural networks as spatio-temporal pattern-forming systems,
  Vol.~61, Rep. Prog. Phys., 1998.

\bibitem{Vilela}
R.~D. Vilela, B.~Lindner, A comperative study of different integrate and fire
  neurons: Spontaneous activity, dynamical response and stimulus indused
  correlation, Phys.Rev.E 80 (2009) 031909.
\newblock \href {http://dx.doi.org/10.1103/PhysRevE.80.031909}
  {\path{doi:10.1103/PhysRevE.80.031909}}.

\bibitem{Lindner}
B.~B.~Lindner, L.~Schimansky-Geier, A.~Longtin, Maximizing spike train
  coherence or incoherence in the leaky integrate-and-fire-model, Phys.Review
  66 (2002) 031916.
\newblock \href {http://dx.doi.org/10.1103/PhysRevE.66.031916}
  {\path{doi:10.1103/PhysRevE.66.031916}}.

\bibitem{Orhan}
E.~Orhan, http://eorhan\@bcs.rochester.edu.

\bibitem{Kouvaris}
N.~Kouvaris, M\, Ensembles of excitable two state units with delayed feedback,
  Phys.Rev.E 82 (2010) 061124.
\newblock \href {http://dx.doi.org/10.1103/PhysRevE.82.061124}
  {\path{doi:10.1103/PhysRevE.82.061124}}.

\bibitem{Tattini}
L.~Tattini, S.~Olmi, A.~Torcini, Coherent periodic activity in excitatory
  erdos-renyi neural networks.the role of network connectivity, Chaos 22 (2012)
  023133.
\newblock \href {http://dx.doi.org/10.1063/1.4723839}
  {\path{doi:10.1063/1.4723839}}.

\bibitem{Luciolii}
S.~Lucioli, A.~Politi, Irregular collective behavior of heterogeneous neural
  networks, Phys.Rev.E 105 (2010) 158104.
\newblock \href {http://dx.doi.org/10.1103/PhysRevLett.105.158104}
  {\path{doi:10.1103/PhysRevLett.105.158104}}.

\bibitem{Zare}
M.~Zare, P.~Grigolini, Cooperation in neural systems: Bridging complexity and
  periodicity, Phys.Rev.E 86 (2012) 051918.
\newblock \href {http://dx.doi.org/10.1103/PhysRevE.86.051918}
  {\path{doi:10.1103/PhysRevE.86.051918}}.

\bibitem{Olmi}
S.~Olmi, A.~Politi, A.~Torcini, Collective chaos in pulse-coupled neural
  networks, Europhys.Lett. 92 (2010) 60007.
\newblock \href {http://dx.doi.org/10.1209/0295-5075/92/60007}
  {\path{doi:10.1209/0295-5075/92/60007}}.

\end{thebibliography}

%------------------------------------------------------------------------------
% Index
%\printindex

%------------------------------------------------------------------------------
\end{document}